\documentclass[prd,amsmath,showkeys,showpacs,preprintnumbers,nofootinbib,superscriptaddress,notitlepage,twocolumn]{revtex4-1}

\usepackage{amssymb}
\usepackage{amsmath}
\usepackage{graphicx}
\usepackage{bm}

\newcommand{\INTM}{\int\frac{d^3\bm p}{(2\pi)^3}}
\newcommand{\eref}{\eqref}

\begin{document}

\title{Hot Quark Matter with an Axial Chemical Potential}
\author{Raoul Gatto}%\email{raoul.gatto@unige.ch}
\affiliation{Departement de Physique Theorique,
 Universite de Geneve, CH-1211 Geneve 4, Switzerland.}
\author{Marco Ruggieri}%\email{ruggieri@yukawa.kyoto-u.ac.jp}
\affiliation{Department of Physics and Astronomy, University of Catania, Via S. Sofia 64, I-95125 Catania}

\begin{abstract}
We analyze the phase diagram of hot quark matter in presence of an
axial chemical potential, $\mu_5$. The latter is introduced to
mimic the chirality transitions induced, in hot Quantum
Chromodynamics, by the strong sphaleron configurations. In
particular, we study the curvature of the critical line
at small $\mu_5$, the effects of a finite quark mass and of a
vector interaction. Moreover, we build the mixed phase at the
first order phase transition line, and draw the phase diagram in
the chiral density and temperature plane. We finally
compute the full topological susceptibility in presence of a
background of topological charge.
\end{abstract}

%\preprint{YITP-11-XY}

\maketitle

\section{Introduction}
Understanding the properties of strongly interacting matter, 
in extreme conditions of high temperature and/or large baryon density, 
is very important to get a deeper knowledge of our universe, 
at the macroscopic as well as at the microscopic levels. 
For example, very hot matter, with estimated temperature of the order of $10^{12}$ Kelvin, 
is produced in heavy ion collisions (HICs) experiments immediately after the collision, 
see for example~\cite{Adare:2009qk} for an indirect measurement of such a temperature 
by the PHENIX collaboration at Brookhaven National Laboratory (BNL). 
As a consequence, it is crucial to make theoretical investigations 
on the phase structure of Quantum Chromodynamics (QCD), 
the theory of strong interactions, in conditions of high temperature, 
in order to give proper interpretation of experimental data, 
at the same time suggesting new phenomena to look for.

The theoretical knowledge on the thermodynamics of QCD at zero baryon density is
fixed by Lattice simulations, which allow first principle numerical computations of 
the relevant thermodynamical quantities. 
In fact, Lattice simulations performed by independent groups show that in the range 
of temperature $(140,180)$ MeV, a crossover from the hadron phase to the deconfinement phase takes place, 
see~\cite{Aoki:2006we,Cheng:2006qk} and references therein. 
On the other hand,
at finite baryon chemical potential it is not possible to use first principle
Lattice simulations to investigate the structure of QCD (with three colors), 
because of the sign problem, see~\cite{deForcrand:2010ys} for a review.
Lattice simulations in the strong coupling limit seem promising 
to avoid the sign problem, 
see~\cite{Nakano:2010bg,deForcrand:2009dh}, even if the problem of the continuum limit within this 
approach has to be considered with care; in particular,
the continuum limit is not reached in~\cite{Nakano:2010bg}, but the
$1/g^2$ evolution in a range of $\beta=6 N_c/g^2 \leq 4 \equiv \beta_s$
is consistent with Lattice Montecarlo simulations, 
for which the continuum limit
is realized for $\beta \approx 6$, hence not too far from $\beta_s$.

The lack of feasible first principle calculations of the QCD thermodynamic properties
at finite baryon chemical potential invokes the use of other theoretical strategies
to investigate the structure of QCD in this regime. The main strategy is the use of some model. 
Among the several models, the Nambu-Jona-Lasinio (NJL) model~\cite{Nambu:1961tp} is very popular,
see~\cite{revNJL} for reviews. In the NJL model, the QCD gluon-mediated interactions are
replaced by effective interactions among quarks, which are built
in order to respect the global symmetries of QCD. Under some
approximations, it is possible to derive the NJL model effective
interaction kernel from first principles QCD,
see~\cite{Kondo:2010ts,Frasca:2011bd}. 
The common feature of the effective models 
is that they share (most of) the symmetries of the QCD Lagrangian. 
At the same time, they are able to describe spontaneous chiral symmetry breaking, 
as well as other kinds of breaking patterns expected at large chemical potential, 
in a self-consistent way. 
Moreover, it has been shown how it is possible to extend the chiral models in order 
to compute quantities which are sensible to confinemt properties of 
a given phase~\cite{Fukushima:2003fw},
see~\cite{Roessner:2006xn,Fukushima:2008wg,Abuki:2008nm,
Kashiwa:2007hw,Herbst:2010rf,Skokov:2010sf,Kahara:2008yg,Sakai:2011gs,Kouno:2011vu,
Sakai:2010rp,Skokov:2010uh,Andersen:2011pr,Sakai:2011fa,Megias:2006bn,Megias:2004hj} 
for recent studies. 
One of the advantage of these models is that they allow for analytic or semi-analytic 
computations at zero as well as at finite chemical potential,
at least at the one-loop level; 
thus they do not rely (in general) on simulations, and are not affected by the 
complex value of the quark determinant, allowing speculations 
about the structure of QCD in regimes where first principles calculations are not feasible.
After standard bosonization procedure, the chiral and deconfinement
transitions are described in terms of collective fields which 
take some expectation value, and whose quantum fluctuations describe
physical particles ($\sigma$ and $\pi$ mesons in the simplest version
of the model).

It has been suggested that very strong magnetic fields are produced during the
very first moments of a noncentral heavy ion collision~\cite{Kharzeev:2007jp,Skokov:2009qp,Voronyuk:2011jd}; 
this has motivated several studies
about the effect of a strong magnetic background on the QCD phase structure, 
see~\cite{D'Elia:2010nq,D'Elia:2011zu,Buividovich:2008wf,Mizher:2010zb,Gatto:2010qs,Agasian:2008tb,
Fraga:2008qn,Chernodub:2010qx,Braguta:2011hq,Agasian:2011st,Callebaut:2011ab,Fukushima:2011nu} 
and references therein.
Moreover, since the temperature of the fireball produced by the collision is
very high, a copious production of topological gluon configurations (i.e., the QCD sphalerons) 
with finite winding number is expected~\cite{Moore:2010jd}, which
induce locally chirality imbalance in the hot plasma as a natural consequence
of the QCD Ward identity. 
The combined effect of the latter and of the magnetic field induces an electric
current along the direction of the magnetic field. This effect, called the Chiral
Magnetic Effect (CME)~\cite{Kharzeev:2007jp,Fukushima:2008xe}, leads to the event-by-event
separation of electric charges with respect to the reaction plane,
which is a parity $({\cal P})$ as well as a ${\cal CP}$-odd effect. 
Experimental data obtained by the STAR collaboration at BNL seem to point
towards the direction of charge separation in collisions~\cite{:2009uh}, even if the
interpretation of such data in terms of the CME is still under debate~\cite{Voronyuk:2011jd,Fukushima:2009ft,Muller:2010jd}.

Because of the expected production of chirality imbalance
in the quark-gluon-plasma phase of QCD, and because of its potential relevance
for the physics of heavy ion collisions, it is of interest to study how
chirality modifies the structure of QCD itself. This is the main scope of
this article. In particular, we continue the study 
of~\cite{Fukushima:2010fe,Chernodub:2011fr,Ruggieri:2011xc,Bayona:2011ab}
in which chirality was induced by an axial chemical potential, $\mu_5$,
conjugated to chirality. Besides the applications to the CME, which are very interesting
on their own because of the potential relevance for the phenomenology of heavy ion collisions, 
the theory at finite $\mu_5$ is interesting because it does not suffer from the sign problem;
as a consequence, grandcanonical ensembles at finite $\mu_5$ can be simulated
on the Lattice~\cite{Yamamoto:2011gk,Yamamoto:2011zx}. These studies might be helpful
to understand the structure of the QCD phase diagram at finite baryon chemical potential.
For example, in~\cite{Ruggieri:2011xc} it has been suggested that the critical
endpoint of the QCD phase diagram might be detected in Lattice simulations
at finite $\mu_5$.
The details about the theoretical framework
will be given in Section II of the article. Here, we wish to stress
the novelties embraced by our study. 

Firstly, we investigate the effect
of the vector interaction, as well as the finite current quark mass, on the
location of the critical endpoint. Such aspects should be taken
into account if a comparison with the Lattice data is desirable, but they have not
been studied in~\cite{Ruggieri:2011xc}.  Moreover, it is of interest to
analyze analytically the effect of the axial chemical potential
on the chiral condensate, and on the curvature of the critical line for restoration
of chiral symmetry. The latter aspect is quite interesting, since it shows
how a competition between the vacuum term and the thermal excitations compete
and eventually lead to a lowering of the critical temperature. This reduction
of the critical temperature was found numerically in the previous model
studies~\cite{Fukushima:2010fe,Chernodub:2011fr,Ruggieri:2011xc} but it was
not investigated in detail. Thirdly, we are interested to the phase diagram
in the $n_5 - T$ plane. This might be interesting because $n_5$ is connected
to the topological charge density, because of the integrated Ward identity.
With an abuse of nomenclature,
we call the latter as canonical ensemble formulation,
in analogy with the case of QCD at finite baryon density. 
Finally, we compute the full topological susceptibility
at zero as well as at finite $\mu_5$, as a function of temperature.
The latter part of our investigation has some overlap
with the model study of Ref.~\cite{Sakai:2011gs}, where
topological susceptibility with a background of topological charge is
computed. In the case of Ref.~\cite{Sakai:2011gs}, the background topological
charge is introduced by adding a $\theta-$term to the QCD action.

The plan of the paper is as follows. In Section II we introduce the 
axial chemical potential. In Section III we summarize the model we use in our
calculations. In Section IV we perform a perturbative analysis at small $\mu_5$
at zero as well as at finite temperature, computing the dependence of the chiral
condensate and of the critical temperature on the axial chemical potential. 
In Section V we relate the phase diagrams in the canonical and grand-canonical ensembles,
and build explicitly the mixed phase. In Section VI we study the effect of
the bare quark mass, and of a vector interaction, on the location of the critical
endpoint of the phase diagram in the $\mu_5 - T$ plane. Finally, we present our conclusions
in Section VII.

\section{The axial chemical potential}
In this Section we define the axial chemical potential, $\mu_5$. It has been already
discussed in several references, 
see~\cite{Ruggieri:2011xc,Fukushima:2008xe,Fukushima:2010fe,Chernodub:2011fr,McLerran:1990de} and references
therein. Therefore we limit ourselves to the basic definitions and to fix our notation.
First of all we introduce the chirality, $N_5 = N_R - N_L$, as the imbalance
between right- and left-handed quarks. In QCD, change of chirality
at zero as well as at finite temperature can be related directly
to the topology of nonperturbative gluon configurations with a
finite winding number, $Q_W$, via the integrated Ward identity,
\begin{equation}
N_5 = n_5 V = 2 N_f Q_W~, \label{eq:Ward}
\end{equation}
where $N_f$ is the number of flavors considered, and we have assumed
that before the interaction with the gluon configuration,
chirality was zero. In Equation~\eqref{eq:Ward} we have introduced
the chiral density $n_5$. The above equation is a consequence of
the strong chiral anomaly; $Q_W$ is related to the topological
charge of the given gluon configuration,
\begin{equation}
Q_W = \frac{g^2}{32\pi^2} \int d^4 x F_{a}^{\mu\nu}
\tilde{F}_{\mu\nu}^a~.
\end{equation}
At high temperature, a copious
production of gluon configurations with nonvanishing winding
number is expected, see for example~\cite{Moore:2010jd} and
references therein. These are the strong (that is, QCD)
sphalerons. Therefore, because of the existence of QCD sphalerons,
chirality can be produced in the high temperature phase of QCD.

The simplest way to treat quark matter with net chirality in
effective models, is
to introduce an axial chemical potential, $\mu_5$, conjugated to
chiral density~\cite{Ruggieri:2011xc,Fukushima:2008xe,Fukushima:2010fe,Chernodub:2011fr}. At the lagrangian level, this amounts to add the
chiral density operator,
\begin{equation}
\mu_5 \bar\psi\gamma^0\gamma^5\psi~,
\end{equation}
to the lagrangian density. 
This procedure is similar to the one
usually adopted to study systems at finite baryon density: in the
latter case, we introduce a baryon chemical potential, $\mu$, that
induces a net baryon density, $n$. 
For example, in the very high
temperature phase of two flavor QCD, the following relation
holds~\cite{Fukushima:2008xe}:
\begin{equation}
n_5 = \frac{\mu_5^3}{3\pi^2} + \frac{\mu_5 T^2}{3}~,
\end{equation}
which shows how a chiral chemical potential induces a net
chirality in the system. In the general case, one needs to compute the relation 
between $n_5$ and $\mu_5$ numerically, 
see~\cite{Fukushima:2010fe,Chernodub:2011fr} and Section V of this article. 

Before going ahead, it is necessary to make few remarks. Firstly, we are aware
that the axial chemical potential cannot be considered as a true chemical potential.
As a matter of fact, $\mu_5$ is conjugated to $n_5$, or to the topological
charge because of the Ward identity. The latter is a not conserved 
quantity in QCD because of the quantum anomaly. In the common
picture of the QCD vacuum with $\mu_5 = 0$, the vanishing average value
of the topological charge is understood as the formation of several local
domains, each one characterized by a finite topological charge; 
the probability to create a domain with charge $Q_W$ is the same
as the probability to create a domain with charge $-Q_W$. As a consequence,
in a volume much larger than the typical domain size, the average value
of the topological charge sums up to zero. However,
flucuations of the topological charge can change the value of the charge
of any local domain. The formal role of the axial chemical potential
is to reproduce the local domains of the QCD vacuum with a net topological
charge as equilibrium states. Because of the fluctuations of the topological charge,
this description is meaningful as long as the time scale is shorter than the
inverse of the topological changing transition rate~\cite{Yamamoto:2011gk}.
This interesting theoretical question has been analyzed also in~\cite{Rubakov:2010qi},
where it is claimed that a proper combination of the chiral density and of
a Chern-Simon term is a conserved quantity, and it is precisely this quantity 
that should be coupled to the axial chemical potential.
We will devote a more detailed study of this problem to a next article.

Another remark is that the theory with $\mu_5 \neq 0$ is a sign-free theory.
It is well known that QCD with three colors suffers the sign problem: namely,
the fermion determinant of QCD with three colors is complex at finite 
quark chemical potential,
making the usual Montecarlo sampling of configurations in the Lattice simulations
not possible when the quark chemical potential is larger than the
temperature (see~\cite{deForcrand:2010ys} for a review). On the other hand,
the theory at finite $\mu_5$ does not suffer the sign problem. As a matter
of fact, $\gamma_5 D(\mu_5) \gamma_5 = D^\dagger (\mu_5)$, where $D$ corresponds
to the Dirac operator with $\mu_5 \neq 0$. 
As a consequence, the fermion determinant is real and positive at $\mu_5 \neq 0$,
and grand canonical ensembles with finite $\mu_5$ can be simulated on the 
Lattice~\cite{Fukushima:2008xe}. Indeed, some Lattice simulation at $\mu_5 \neq 0$ 
has been already performed to study the chiral magnetic effect on the Lattice,
see~\cite{Yamamoto:2011gk,Yamamoto:2011zx}. In these references, a preliminary study of
the phase structure in the $\mu_5 - T$ plane is also addressed.

\section{The model}
In this Section, we describe the model that we use in
calculations, namely the Nambu-Jona-Lasinio model improved with
the Polyakov loop (PNJL in the following)~\cite{Fukushima:2003fw}.
In the PNJL model, quark propagation in the medium is described by
the following lagrangian density:
\begin{equation}
{\cal L} = \bar q\left(i\gamma^\mu D_\mu - m\right)q +G\left[\left(\bar qq\right)^2 + \left(i\bar
q\gamma_5\bm\tau q\right)^2\right]~;\label{eq:1ooo}
\end{equation}
here $q$ is the quark Dirac spinor in the fundamental
representation of the flavor $SU(2)$ and the color group;
$\bm\tau$ correspond to the Pauli matrices in flavor space. A sum
over color and flavor is understood. The covariant derivative
embeds the QCD coupling with the background gluon field which is
related to the Polyakov loop, see below.

The PNJL model has access to the
expectation value of the Polyakov loop, $L$,
which is sensible to confinement or deconfinement properties of a
given phase. In order to compute $\langle L \rangle$ we introduce a static,
homogeneous and Euclidean background temporal gluon field, $A_0 =
iA_4 = i \lambda_a A_4^a$, coupled minimally to the quarks via the
QCD covariant derivative. Then
$W = \text{Tr}_c\exp\left(i\beta\lambda_a A_4^a\right)$ and $L=W/3$,
where $\beta = 1/T$. In the Polyakov gauge, which is convenient
for this study, $A_0 = i\lambda_3 \phi + i \lambda_8 \phi^8$;
moreover, for simplicity we take $L = L^\dagger$ from the
beginning as in~\cite{Roessner:2006xn}, which implies $A_4^8 = 0$.

In our computation we follow the idea implemented
in~\cite{Sakai:2010rp}, which brings to a Polyakov-loop-dependent
coupling constant:
\begin{equation}
G = g\left[1 - \alpha_1 L L^\dagger -\alpha_2(L^3 +
(L^\dagger)^3)\right]~,\label{eq:Run}
\end{equation}
The ansatz in the above equation was inspired
by~\cite{Kondo:2010ts} in which it was shown
explicitly that the NJL vertex can be derived in the infrared
limit of QCD, it has a non-local structure, and it acquires a
non-trivial dependence on the phase of the Polyakov loop. This
idea has been analyzed recently in~\cite{Braun:2011fw}, where the
effect of the confinement order parameter on the four-fermion
interactions and their renormalization-group fixed-point structure
has been investigated. The numerical values of $\alpha_1$ and
$\alpha_2$ have been fixed in~\cite{Sakai:2010rp} by a best fit of
the available Lattice data at zero and imaginary chemical
potential of Ref.~\cite{D'Elia:2009qz,Bonati:2010gi}. In
particular, the fitted data are the critical temperature at zero
chemical potential, and the dependence of the Roberge-Weiss
endpoint on the bare quark mass. The best fit procedure leads to
$\alpha_1 = \alpha_2 \equiv \alpha = 0.2$, which are the values we adopt in 
this article.

In the one-loop approximation, the effective potential of this
model is given by~\cite{Ruggieri:2011xc}
\begin{eqnarray}
V &=& {\cal U}(L,L^\dagger,T) + \sigma^2 G   -N_c N_f\sum_{s=\pm 1}\INTM \omega_s \nonumber \\
&&-\frac{N_f}{\beta}\sum_{s=\pm 1}\INTM\log\left(F_+ F_-\right)~, \nonumber \\
&&~\label{eq:OB}
\end{eqnarray}
where $\sigma = \bar q q$ is a scalar collective field representing
the mean field and the quantum fluctuations of the operator $\bar q q$, and $G$ is defined in Eq.~\eqref{eq:Run};
moreover,
\begin{equation}
\omega_s = \sqrt{(|\bm p| s -\mu_5)^2 + m_q^2}~, \label{eq:iii}
\end{equation}
corresponds to the pole of the quark propagator, and $m_q = m
-2\sigma G$ is the constituent quark mass; the index $s$ denotes
the helicity projection. In Eq.~\eqref{eq:OB} we have introduced
the functions
\begin{eqnarray}
F_- &=& 1+3L e^{-\beta(\omega_s - \mu)} +3L^\dagger
e^{-2\beta(\omega_s - \mu)} + e^{-3\beta(\omega_s -
\mu)}~,\nonumber \\
&& \\
F_+ &=& 1+3L^\dagger e^{-\beta(\omega_s + \mu)} +3L
e^{-2\beta(\omega_s + \mu)} + e^{-3\beta(\omega_s +
\mu)}~,\nonumber \\
&&
\end{eqnarray}
which are responsible for the statistical confinement property of
the model at low temperature~\cite{Fukushima:2003fw}.

In right hand side of the first line of Equation~\eqref{eq:OB} the
momentum integral corresponds to the vacuum quark fluctuations
contribution to the thermodynamic potential. We treat the
divergence in this term phenomenologically, introducing a momentum
cutoff, $\Lambda$, in the vacuum term; the numerical value of
$\Lambda$ will be then fixed by requiring that the quark
condensate, the pion mass and the pion decay constant computed in
the model are in agreement with the phenomenological values.
Before going ahead, it is useful to remind that within the quark-meson
model a renormalization procedure is feasible~\cite{Skokov:2010sf}, the model itself
being renormalizable. The renormalizability of the model might 
be useful to remove the cutoff effects that, instead, appear in 
the NJL model (see the discussion in the next Sections).
On the other hand, the renormalization program of the energy
density at finite $\mu_5$ is not trivial, since a nonvanishing
$\mu_5$ induces further divergences in the theory, as it can 
be argued by an inspection of the $\mu_5-$dependence of the
vacuum energy, see for example Eq.~\eqref{eq:OP}. Hence, proper
renormalization conditions should be adopted.

The potential term $\mathcal{U}$ in Eq.~\eqref{eq:OB} is built by
hand in order to reproduce the pure gluonic lattice data with $N_c
= 3$~\cite{Roessner:2006xn}. We adopt the following logarithmic
form,
\begin{equation}
 \begin{split}
 & \mathcal{U}[L,\bar L,T] = T^4\biggl\{-\frac{a(T)}{2}
  \bar L L \\
 &\qquad + b(T)\ln\bigl[ 1-6\bar LL + 4(\bar L^3 + L^3)
  -3(\bar LL)^2 \bigr] \biggr\} \;,
 \end{split}
\label{eq:Poly}
\end{equation}
with three model parameters (one of four is constrained by the
Stefan-Boltzmann limit),
\begin{eqnarray}
 a(T) &=& a_0 + a_1 \left(\frac{T_0}{T}\right)
 + a_2 \left(\frac{T_0}{T}\right)^2 , \\
 b(T) &=& b_3\left(\frac{T_0}{T}\right)^3.
\label{eq:lp}
\end{eqnarray}
The standard choice of the parameters reads $a_0 = 3.51$, $a_1 =
-2.47$, $a_2 = 15.2$ and $b_3 = -1.75$. The parameter $T_0$ in
Eq.~\eqref{eq:Poly} sets the deconfinement scale in the pure gauge
theory. In absence of dynamical fermions one has $T_0 = 270$
\text{MeV}. However, dynamical fermions induce a dependence of
this parameter on the number of active
flavors~\cite{Herbst:2010rf}. For the case of two light flavors to
which we are interested here, we take $T_0 = 190$ MeV as
in~\cite{Sakai:2010rp}. Also for the remaining parameters we
follow~\cite{Sakai:2010rp} and take $\Lambda = 631.5$ MeV, $m=5.5$
MeV and $g=5.498\times 10^{-6}$ MeV$^{-2}$.

We notice that the PNJL model considered here, which is dubbed
Extended-PNJL in~\cite{Sakai:2010rp}, has been tuned in order to
reproduce quantitatively the Lattice QCD thermodynamics at zero
and imaginary quark chemical potential. Hence, it represents a
faithful description of QCD, in terms of collective degrees of
freedom related to chiral symmetry breaking and deconfinement.

\section{Perturbative analysis}
In the case of small $\mu_5$, we can make some analytical and
semianalytical estimate of the effect of the axial chemical
potential on quark condensation, and on the critical line. In this
Section we restrict our analysis to the pure NJL model case,
corresponding to take $L = L^\dagger =1$ in Eq.~\eqref{eq:OB}, and
$\alpha_1 = \alpha_2 = 0$ in Eq.~\eqref{eq:Run}. This simplifies
the numerical analysis. On the other hand, the results obtained
here will not be modified qualitatively by the Polyakov loop,
since the effect of the latter is just a suppression of colored
states below $T_c$. Moreover, we work in the chiral limit; this
simplification allows to define rigorously the chiral phase
transition, and compute unambiguously the critical temperature.

\subsection{Zero temperature: chiral condensate}
To begin with, we consider the zero temperature case, and we
compute the shift of the chiral condensate induced by $\mu_5$,
showing that the chiral chemical potential acts as a catalyzer of
chiral symmetry breaking. In fact, the $\mu_5-$dependent zero
temperature correction to the effective potential is given by
\begin{equation}
V_1^0 = -\frac{N_c N_f}{2\pi^2}\mu_5^2 m_q^2
 F\left(\frac{m_q}{\Lambda}\right)~, \label{eq:OP}
\end{equation}
where we have assumed $\mu_5 \ll m_q$; in Eq.~\eqref{eq:OP} we
have introduced the function
\begin{equation}
F(x) = \log\frac{1+\sqrt{1+x^2}}{x} - \frac{1}{\sqrt{1+x^2}}~.
\label{eq:Fx}
\end{equation}
It is easy to prove that $F(x)$ is always positive, thus making
$V_1^0 <0$. Hence, the energy density of a broken phase and $\mu_5
\neq 0$ is smaller than that of a phase with unbroken symmetry and
the same value of the chiral condensate.

We can use Eq.~\eqref{eq:OP} to analyze the perturbative solution
of the gap equation at zero temperature. As a matter of fact, for
small values of $\mu_5$ we can look for a solution of the gap
equation in the form $m_q = \bar{m}_q + \delta m_q$, where
$\bar{m}_q$ satisfies the gap equation at $\mu_5 = 0$, and $\delta
m_q$ corresponds to the $\mu_5$-dependent contribution. An
elementary computation shows that
\begin{equation}
\delta m_q = \left.-\frac{1}{M_\sigma^2}\left(\frac{\partial
V_1^0}{\partial m_q}\right)\right|_{m_q = \bar{m}_q}~,
\label{eq:GH}
\end{equation}
where we have defined $M_\sigma^2 \equiv \left(\partial^2
V_0/\partial m_q^2\right)_{m_q = \bar{m}_q}$ and $V_0 \equiv
V(\mu_5 = 0)$. By definition, $M_\sigma^2 > 0$, since $\bar{m}_q$
corresponds to a minimum of the effective potential at $\mu_5 =
0$. Moreover, for $m_q \ll \Lambda$ we find $\left(\partial
V_1^0/\partial m_q\right) \propto m_q \log
\left(m_q/\Lambda\right)$ which is negative (we have verified that
the sign does not change as long as $m_q \leq \Lambda$). We
conclude that $\delta m_q > 0$, showing that $\mu_5$ favors the
spontaneous breaking of chiral symmetry at zero temperature.

\subsection{Finite temperature: chiral condensate}
\begin{figure}[t!]
\begin{center}
\includegraphics[width=8.5cm]{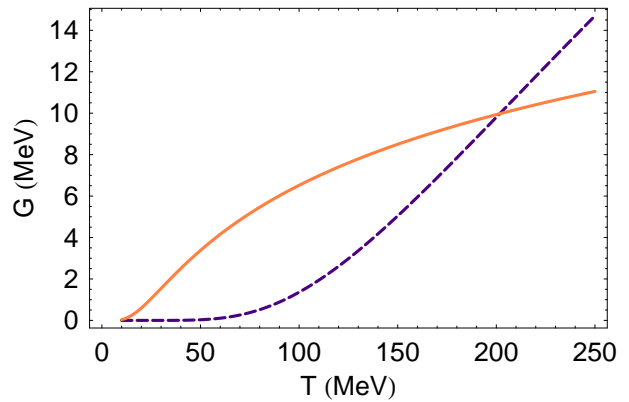}
\caption{Function $G$ of Eq.~\eqref{eq:DT} versus temperature, for
two values of the constituent quark mass, $m_q$. Orange solid line
corresponds to $m_q = 50$ MeV; Indigo dashed line corresponds to
the case $m_q = 350$ MeV.} \label{Fig:GFunc}
\end{center}
\end{figure}

We can extend the analysis of the previous subsection, to the case
of finite temperature. For the perturbative solution of the gap
equation, the derivative of the effective potential with respect
to the quark mass is needed, see Eq.~\eqref{eq:GH}. At finite
temperature, the $\mu_5$-dependent contribution of the effective
potential is $V_1 = V_1^0 + V_1^T$ with $V_1^0$ given by
Eq.~\eqref{eq:OP}. Thus, Eq.~\eqref{eq:GH} is replaced by
\begin{equation}
\delta m_q = \left.-\frac{1}{M_\sigma^2}\left(\frac{\partial
V_1^0}{\partial m_q} + \frac{\partial V_1^T}{\partial
m_q}\right)\right|_{m_q = \bar{m}_q}~. \label{eq:GHt}
\end{equation}
The expression of $V_1^T$ is not informative, thus it is not
necessary to report it here. Its derivative with respect to the
quark mass is more interesting,
\begin{equation}
\frac{\partial V_1^T}{\partial m_q} = N_c N_f \mu_5^2 G(m_q,T)~.
\label{eq:DT}
\end{equation}
In the above equation, $G(m_q,T)$ corresponds to a function
defined in terms of a convergent numerical integral. In
Fig.~\ref{Fig:GFunc}, we plot $G$ as a function of temperature,
for two different values of the quark mass. We find that $G>0$. As
a consequence, because of Eq.~\eqref{eq:GHt}, the
$\mu_5$-dependent contribution of the thermal fluctuations tends
to reduce the value of the constituent quark mass (i.e., of the
chiral condensate).

%The different effect of vacuum and thermal fluctuations on the
%chiral condensate as a function of $\mu_5$, explains why in the
%Quark-Meson model computation of~\cite{Chernodub:2011fr},

\subsection{Critical line}
We compute in this subsection the critical temperature, $T_c$, as
a function of $\mu_5$, and show at the same time that the chiral
phase transition is of the second order. To this end we perform an
expansion of the effective potential near $T_c$.
In fact, the order parameter around $T_c$ is small enough that a
series expansion in powers of $\sigma/T$ is justified.
Since we restrict ourselves to the case of the pure NJL model,
there is no dependence from the Polyakov loop in the NJL coupling constant, $G$.

One can write
\begin{equation}
V  = V_0 + \frac{\alpha_2}{2}G^2\sigma^2 + \frac{\alpha_4}{4}G^4\sigma^4
+ \dots~, \label{eq:GL1}
\end{equation}
where the dots denote higher order terms; $V_0$ is the potential
at $\sigma=0$: it is independent on the condensate, thus it is
just a number which does not affect the physics of the problem. We notice that
because of our definition of the $\sigma$ field, which has the dimension of
a cubic mass, we have extracted the proper 
powers of the NJL coupling constant from the definitions of the coefficients, in order
to give the common mass dimension to the coefficients themselves.
At the second order transition point $T = T_c$ one has $\alpha_2 = 0$
and $\alpha_4 > 0$. Therefore, to determine the critical
temperature as a function of $\mu_5$ it is enough to determine the
zeros of $\alpha_2$ in the $\mu_5-T$ plane. Next, we compute
$\alpha_4$ to check that the transition points correspond to a
second order phase transition.

Computing the second derivative of the effective potential in
Equation~\eref{eq:OB} we find, after some algebra,
\begin{equation}
\alpha_2 = \alpha_{2,0} + \alpha_{2,2}\mu_5^2~,
\end{equation}
where
\begin{eqnarray}
\alpha_{2,0} &=& \frac{2}{G}-\frac{2N_c N_f}{\pi^2}\Lambda^2 +
\frac{2N_c N_f}{3} T^2~, \label{eq:G1}\\
\alpha_{2,2} &=& \frac{4 N_c
N_f}{\pi^2}\left(\log\frac{T}{\Lambda} + c\right)~, \label{eq:G2}
\end{eqnarray}
and the numerical factor $c=1.36$ arises from a convergent
numerical integral.

At small $\mu_5$ the equation $\alpha_2 = 0$, which determines the
critical temperature, can be solved perturbatively: from the
condition $\alpha_{2,0} = 0$ we find
\begin{equation}
T_c^0 = \left(\frac{3 \Lambda^2}{\pi^2} - \frac{3}{N_c N_f
G}\right)^{1/2}~;
\end{equation}
using the parameters of the model we find $T_c^0 = 174$ MeV. Then,
writing $T_c = T_c^0 + \delta T$ and solving $\alpha_{2,0} +
\alpha_{2,2}\mu_5^2 = 0$ for $\delta T$ we find
\begin{equation}
T_c = T_c^0\left(1 -\frac{3\mu_5^2}{4 N_c N_f
(T_c^0)^2}\alpha_{2,2}(T_c^0)\right)~.  \label{eq:TCrr}
\end{equation}
Using the parameters given above we find $\alpha_{2,2}(T_c^0) =
0.17$, which implies that the critical temperature is a decreasing
function of $\mu_5$. Hence, a finite $\mu_5$ favors the disordered
phase at finite temperature.

The last step is to check the sign of $\alpha_4$, to be confident
that the transition is of the second order. Since we make a
perturbative analysis at $\mu_5 \ll T$, we may assume $\mu_5$
infinitesimal of the same order of $\sigma$, hence terms of the
order of ${\cal O}(\mu_5^2 \sigma^4)$ can be regarded as ${\cal
O}(\sigma^6)$ and can be neglected. At the leading order, it is
therefore enough to compute the coefficient at $\mu_5 = 0$,
\begin{equation}
\alpha_4 =  \frac{48 N_c N_f}{\pi^2}\left(\log\frac{\Lambda}{T} -
b\right)~, \label{eq:A4rr}
\end{equation}
where $b=0.90$ arises from a convergent integral computed
numerically. Using the values of $\Lambda$ and $T_c^0$ we find
that $\alpha_4 = 11.35$ at $T = T_c^0$; being it positive, the
phase transition is of the second order.

Before going ahead, it is useful to summarize the results of this
Section. Eq.~\eqref{eq:GHt} represents the correction to the
solution of the gap equation due to $\mu_5 \neq 0$. The first and
second addenda in the right hand side of the equation correspond
to the zero temperature and finite temperature contributions,
respectively. At zero temperature, only the first contribution
survives, and it is negative, leading to a positive shift of the
constituent quark mass. On the other hand, the contribution of
thermal fluctuations is positive. Thus at a given temperature,
there is a competition between the effect of $\mu_5$ on the vacumm
and thermal contributions to the gap equation. In passing, we
notice that if vacuum fluctuations were neglected in the
thermodynamic potential, then the thermal contribution in
Eq.~\eqref{eq:GHt} would lower the constituent quark mass at
finite temperature as $\mu_5$ is switched on. Finally, the results
in Equations~\eref{eq:G2},~\eref{eq:TCrr} and~\eref{eq:A4rr} show
that at zero and small $\mu_5$, the chiral phase transition is of
the second order, and the critical temperature is a decreasing
function of the chiral chemical potential. This scenario for the
critical line is in agreement with the results
of~\cite{Chernodub:2011fr,Ruggieri:2011xc,Fukushima:2010fe}, where
a numerical procedure of minimization of the one-loop potential
has been adopted.

\section{Grandcanonical and Canonical Phase Diagrams}
In this Section, we describe the expected phase structure of QCD
at finite temperature and axial chemical potential, as predicted
by the PNJL model. The grandcanonical phase diagram, already
discussed in~\cite{Ruggieri:2011xc}, corresponds to the map of the
different phases in the $\mu_5 - T$ plane. On the other hand, for
the case of the canonical ensemble phase diagram, $\mu_5$ is
replaced by the chiral density $n_5$, the latter being defined as
\begin{equation}
n_5 = - \frac{\partial\Omega}{\partial\mu_5}~, \label{eq:n5}
\end{equation}
where the derivative has to be computed at the global minimum of
the grand potential.

The computation of $n_5$ in the model needs some care. In particular, 
at $m_q \neq 0$ the divergence of the vacuum energy at finite $\mu_5$
is transmitted to $n_5$. 
As a matter
of fact, $n_5$ can be formally split into a vacuum and a
thermal parts. The thermal part is convergent, and its derivative
is finite as well, therefore it does not lead to any difficulty.
On the other hand, the $T=0$ contribution is divergent, and its
derivative with respect to $\mu_5$ is divergent as well in the
case $m_q\neq0$. This can be realized easily if we take for a moment
the limit $\Lambda \gg \mu_5,m_q$, in which we find
\begin{equation}
n_5 = \frac{N_c N_f}{2\pi^2}\left[\frac{2}{3}\mu_5^3   +
2m_q^2\mu_5\left(\log\frac{2\Lambda}{m_q} - 1\right) \right]~.
\label{eq:Pot}\end{equation}
In the case $m_q=0$ we obtain the result of~\cite{Fukushima:2008xe},
which is not affected by the ultarviolet divergence.
Because of the latter, we cannot take the limit
$\Lambda\rightarrow\infty$ in the calculation of the
phase space integrals which are involved in the expressions
of $n_5$. Therefore
in the numerical computation of $n_5$, we cutoff the vacuum contribution at
$p=\Lambda$ unless otherwise stated, for internal consistency.

Before the discussion of the results, it is useful to comment
briefly on the expected effect of the finite cutoff on $n_5$. 
If $\mu_5\gg m_0$, which is appropriate in the quark-gluon plasma phase, 
we can neglect the mass term in the vacuum
contribution of the fermion determinant to $n_5$, which we call ${\cal C}_0$:
\begin{equation}
{\cal C}_0=\frac{N_c N_f}{3\pi^2}\left[ \mu_5^3\theta(\Lambda-\mu_5) +
\Lambda^3\theta(\mu_5- \Lambda) \right]~. \label{eq:func}
\end{equation}
We notice that the term proportional to $\Lambda^3$ in the
above equation does not appear in Eq.~\eqref{eq:Pot} since 
in the latter we have taken the limit $\Lambda\gg\mu_5$.
As soon as $\mu_5 >\Lambda$, the vacuum contribution to the 
chiral density saturates because of the existence of a cutoff in the theory, 
as shown by Eq.~\eqref{eq:func}.
In the numerical calculations, we expect that the saturation effect
will be softened by the thermal fluctuation contributions.  As a consequence we
expect to measure some mild saturation effect as $\mu_5 \approx \Lambda$. This saturation 
is observed on the Lattice as well, see~\cite{Yamamoto:2011gk,Yamamoto:2011zx}.
In last analysis, this is not a serious trouble, since we expect the model
to be consistent only when masses are smaller than the cutoff, in
this case for regime $\mu_5 < \Lambda$.

\begin{figure}[t!]
\begin{center}
\includegraphics[width=8.5cm]{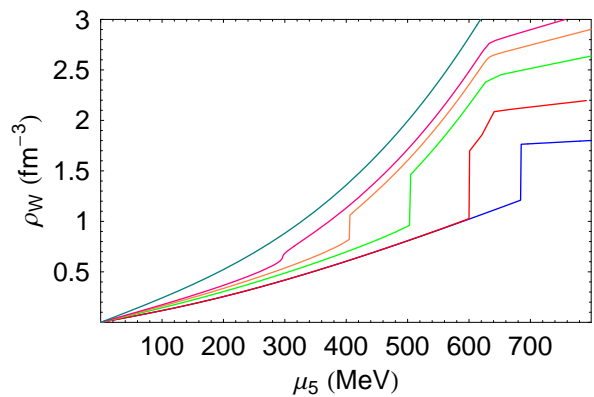}
\caption{Normalized chiral density, $\rho_W\equiv n_5/2N_f$, as a function of the chiral
chemical potential, at several values of the temperature. From
right to left, temperature is equal to $0.4 T_c$, $0.6 T_c$, $0.8
T_c$, $0.9 T_c$, $T_{c5}$, $1.1 T_c$ respectively. }
\label{Fig:data}
\end{center}
\end{figure}

In Fig.~\ref{Fig:data} we plot the normalized chiral density, 
$\rho_W \equiv n_5/2N_f$, as a function of $\mu_5$
for several values of the temperature. In the Figure, $T_{c} =
173.9$ MeV corresponds to the critical temperature for
deconfinement and chiral symmetry restoration at $\mu_5 = 0$. At
large values of $\mu_5$ we measure the expected saturation of
$\rho_W$. This saturation is not physical but a mere artifact of
our regularization scheme. Hence we will consider trustable only
the results in the range $\mu_5 < \Lambda$; when the results
outside the above specified range are shown, it is done only for
completeness.

\begin{figure}[t!]
\begin{center}
\includegraphics[width=8.5cm]{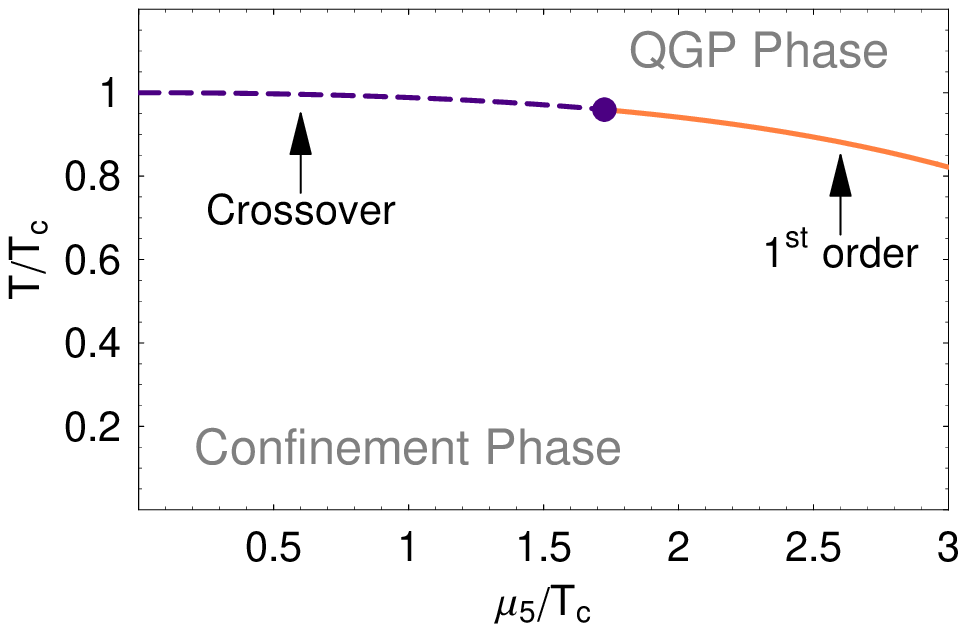}\\
\vspace{0.3cm}
\includegraphics[width=8.5cm]{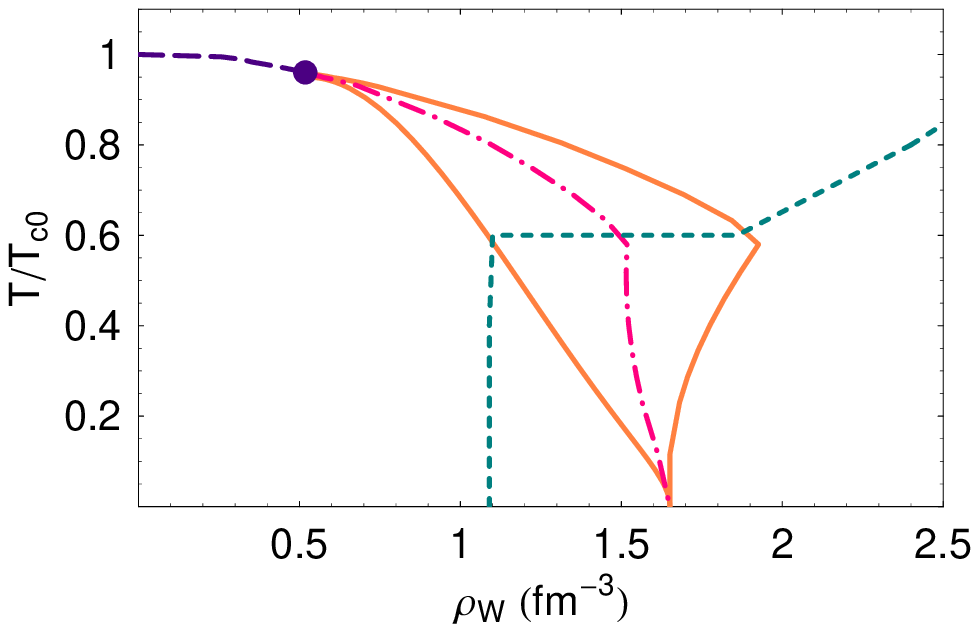}
\caption{(color online). {\em Upper panel:} phase diagram in the
grandcanonical ensemble. The indigo dashed line corresponds to the
chiral and deconfinement crossover; the solid line denotes the
first order phase transition. The indigo dot is the critical
endpoint. The scale $T_c = 173.9$ MeV corresponds to the critical
temperature at $\mu_5 = 0$. {\em Lower panel:} phase diagram in
the temperature-chiral density plane. Solid lines
correspond to the boundaries of the mixed phase. On the left line
the volume fraction, $c$, of hadron phase is $c=1$; on the right
line we find $c=0$. Dashed line denotes the chiral and
deconfinement crossovers. Dot-dashed line corresponds to the
values of $T(\rho_W)$ at which $c=1/2$. In the portion of the
phase diagram below the dotted line we find $\mu_5
> \Lambda$. Finally, the indigo dot denotes the critical
endpoint.} \label{Fig:pdQ}
\end{center}
\end{figure}

Next we turn to the phase structure. The grandcanonical phase
diagram has been computed in~\cite{Ruggieri:2011xc}. It is useful,
however, to briefly summarize here the results
of~\cite{Ruggieri:2011xc}, in order to facilitate the comparison
with the canonical phase diagram. In the upper panel of
Fig~\ref{Fig:pdQ} we plot the phase diagram in the $\mu_5 - T$
plane. The indigo dashed line corresponds to the chiral and
deconfinement crossover; the solid line denotes the first order
phase transition. The indigo dot is the critical endpoint. The
scale $T_c = 173.9$ MeV corresponds to the critical temperature at
$\mu_5 = 0$.

The critical line in the phase diagram in Fig.~\ref{Fig:pdQ} is identified with the
peak of the derivative $dL/dT$. Within few MeV, we have found
that the latter coincides with the location of the peak of
$|d\sigma/dT|$, for the values of $\mu_5$ analyzed in this work. 
Thus, within the model ad hand, the deconfinement and the chiral
symmetry restoration take place simultaneously. In last analysis,
this is related to the explicit dependence of the NJL coupling
on the Polyakov loop, and to the mechanism that leads
to deconfinement in the model. As a matter of fact, at finite temperature
the one- and two-quark contributions to the grand potential
act as efficient $Z_3-$breaking terms, which in turn
favor the $L\neq0$ state. As a consequence, the effective
value of the NJL coupling which is responsible for the
spontaneous chiral symmetry breaking is reduced with respect to the zero
temperature case, as soon as $L\neq0$, see Eq.~\eqref{eq:Run}.
Thus, it is expected that deconfinement in the extended PNJL model 
implies chiral symmetry restoration, as it is confirmed by the numerical
results. The inverse assertion, namely that chiral symmetry restoration implies
deconfinement, is not true and indeed it is not necessarily realized
in the model, unless an explicit backreaction on the Polyakov
loop effective potential is introduced, see for 
example~\cite{Abuki:2008nm,Herbst:2010rf,Sakai:2011fa}.    

The grandcanonical phase diagram can be translated to a canonical
one, by replacing $\mu_5$ with $\rho_W$. This program is easily accomplished
once we use Eq.~\eqref{eq:n5} to compute
$\rho_W$, once the physical values of $\langle\bar q q\rangle$ and
$L$ are known. As already explained in the Introduction,
the use of the term canonical has to be taken as an abuse
of nomenclature. Indeed $n_5$ is not a conserved
quantity, because of the fluctuations of the topological charge
in QCD and the Ward identity. Thus, discussing about $n_5$
is meaningful only if the time of observation of the system
is smaller than the typical time needed to observe a fluctuation
of the topological charge, the latter being related to the inverse
of the sphaleron transition rate.

In the lower panel of Fig.~\ref{Fig:pdQ} we plot
the phase diagram in the $\rho_W - T$ plane. In the figure, the
indigo dashed line denotes the chiral and deconfinement crossovers
(the two coincide within numerical uncertainties within this model
calculation). The indigo dot denotes the critical endpoint. Its
coordinates in the phase diagram are
\begin{equation}
\rho_{5}^c = 0.518~\text{fm}^{-3}~,~~~T_{5}^c = 167~\text{MeV}~.
\end{equation}
The orange solid lines correspond to the boundaries of the mixed
phase, which develops in correspondence of the first order phase
transition line of the grancanonical phase diagram. The
composition of the mixed phase at the temperature $T$ can be
evaluated easily as follows. Given the total charge density
$\rho_W$, then the volume fractions of the ordered, $c$, and
disordered, $d$, phases have to satisfy $c \rho_1 + d \rho_2 =
\rho_W$ with $c+d=1$; here $\rho_{1,2}$ denote the values of the
topological charge density on the two boundaries of the mixed
phase region. It follows then
\begin{equation}
c = \frac{\rho_2 - \rho_W}{\rho_2 -\rho_1}~; \label{eq:9}
\end{equation}
on the left line the volume fraction $c=1$, corresponding to a
homogeneous pure ordered phase; on the right line $c=0$,
corresponding to a pure disordered phase. For reference, in
Fig.~\ref{Fig:pdQ} we plot a pink dot-dashed line which
corresponds to the values of $T(\rho_W)$ at which $c=1/2$.
Finally, in the portion of the phase diagram below the green
dotted line we find $\mu_5 > \Lambda$, in which cutoff artifacts
are relevant. At small temperatures the green line is almost
vertical, because the topological charge density turns out to be
very insensitive of temperature, see Fig.~\ref{Fig:data}.

\section{Influence of the quark mass and of the vector interaction}
In this Section, we wish to investigate the quantitative effect
of a nonphysical bare quark mass, as well as of the presence of the vector
interaction, on the critical line in the $\mu_5 - T$ plane. In particular,
we wish to compute the evolution of the critical endpoint coordinates
as a function of the bare quark mass, and of the coupling strength in
the vector channel.

\subsection{The effect of the bare quark mass}
In this Section, we study the effects of the bare quark mass on
the location of the critical endpoint CP$_5$. Estimation of the
effect of the bare quark mass is very important, because Lattice
simulations at $\mu_5 \neq 0$ nowadays are performed with a
numerical value of the quark mass which is larger than the
physical value. In Lattice simulations, the non physical value of
the quark mass is reflected into a non physical value of the pion
mass. In~\cite{Yamamoto:2011gk,Yamamoto:2011zx}, the value of the
pion mass in the vacuum is $m_\pi \approx 400$ MeV.

\begin{figure}[t!]
\begin{center}
\includegraphics[width=8.5cm]{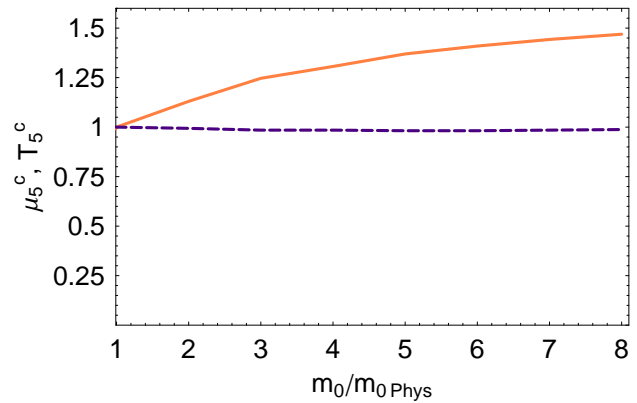}
\caption{({\em Color online}). Evolution of the critical endpoint
coordinates as a function of the bare quark mass. Solid orange
line corresponds to the critical value of the axial chemical
potential; dashed indigo line corresponds to the critical endpoint
temperature. Axial chemical potential and temperature are measured
in units of the endpoint coordinates at $m_0 =
m_{0,\text{Phys}}$.} \label{Fig:M0}
\end{center}
\end{figure}

In Fig.~\ref{Fig:M0} we plot the critical endpoint coordinates as
a function of the bare quark mass. In the figure, the solid orange
line corresponds to the critical value of the axial chemical
potential; the dashed indigo line corresponds to the critical
endpoint temperature. Axial chemical potential and temperature are
measured in units of the endpoint coordinates at $m_0 =
m_{0,\text{Phys}}$ with $m_{0,\text{Phys}} = 5.5$ MeV, which
corresponds to $m_\pi\approx 139$ MeV. For comparison, at the
largest value of quark mass considered here we find $m_\pi \approx
400$ MeV.

Our results show that the critical endpoint temperature, $T_5^c$,
is not so much affected by the quark mass. As we will show in the
next Section, this inertia of $T_5^c$ is also present when we
switch on a vector interaction. On the other hand, the critical
endpoint axial chemical potential, $\mu_5^c$, is strongly affected
by the quark mass. In more detail, the larger value of $m_0$, the
larger $\mu_5^c$. This is quite easy to understand naively, since
the bare quark mass turns the chiral transition to a crossover. At
$\mu_5=0$, the chiral crossover is smoothed as $m_0$ is increased.
Therefore, it is natural to expect that a larger value of $\mu_5$
is necessary to experience a first order phase transition, if $m_0
> m_{0,{\text{Phys}}}$. This might partly explain why the critical endpoint CP$_5$
is not yet detected in Lattice
simulations~\cite{Yamamoto:2011gk,Yamamoto:2011zx}.

\subsection{The role of a vector interaction}
In this Section we briefly comment on the role of a vector
interaction on the phase structure of the model. To this end, we
add to the lagrangian density the term
\begin{equation}
{\cal L}_V = -G_V\left[\left(\bar\psi\gamma^\mu\psi\right)^2 +
\left(\bar\psi\gamma^\mu\gamma_5\psi\right)^2 \right]~;
\label{eq:GV}
\end{equation}
we do not include the interaction in the triplet channel, since at
the one-loop level it gives rise to terms which couple the isospin
density, $\delta n= n_d - n_u$, to the isospin density operator;
these are not relevant as long as we do not introduce the isospin
chemical potential.

At the mean field level one has
\begin{eqnarray}
{\cal L}_V &=& G_V n^2 - 2 G_V n \bar\psi\gamma^0\psi \nonumber
\\
&& + G_V n_5^2 - 2 G_V n_5 \bar\psi\gamma^0\gamma^5\psi~;
\end{eqnarray}
as usual, we have defined $n = \langle\psi^\dagger\psi\rangle$ and
$n_5 = \langle\psi^\dagger\gamma^5\psi\rangle$. Thus at the mean
field level, the chemical potentials for quark number density and
for chiral density are shifted because of the vector interaction,
\begin{eqnarray}
\mu &\rightarrow& \mu - 2 G_V n~, \label{eq:POI}\\
\mu_5 &\rightarrow& \mu_5 - 2 G_V n_5~.  \label{eq:POI5}
\end{eqnarray}

The previous equations are useful to grasp the effect of $G_V \neq
0$ on the phase structure. As a matter of fact, if $G_V > 0$ then
quarks propagate in a thermal bath at an affective chemical
potential which is smaller than the true chemical potential,
because of Eq.~\eqref{eq:POI}. As a consequence, the transition at
finite $\mu$ is smoothed, and the critical endpoint moves to
higher values of the chemical potential. On the same footing,
because of Eq.~\eqref{eq:POI5}, the effective chiral chemical
potential is smaller than $\mu_5$. Therefore we expect that the
critical endpoint coordinate moves to higher values of $\mu_5$
compared to the case $G_V = 0$.

This reasonings are confirmed by our concrete numerical
computations. In the case of $G_V \neq 0$, the value of $n_5$ has
to be computed self-consistently by means of the number equation~\eqref{eq:n5},
which has to be solved, at any given value of $\mu_5$ and $T$,
together with the gap equations,
\begin{equation}
\frac{\partial\Omega}{\partial\sigma} =
\frac{\partial\Omega}{\partial L} = 0~. \label{eq:GAP}
\end{equation}

\begin{figure}[t!]
\begin{center}
\includegraphics[width=8.5cm]{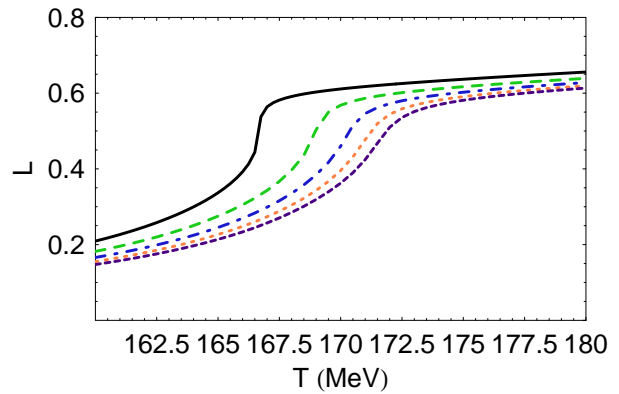}
\caption{({\em Color online}). Expectation value of the Polyakov
loop in the pseudocritical range as a function of temperature, for
several values of the strength of the coupling in the vector
channel, at $\mu_5 = 300$ MeV. Black solid line corresponds to
$G_V=0$. Green dashed line corresponds to $G_V = 0.25 G$. Blue
dot-dashed line corresponds to $G_V = 0.5 G$. Orange dotted line
corresponds to $G_V = 0.75 G$. Finally, indigo short dashed line
corresponds to $G_V = G$. } \label{Fig:cotro}
\end{center}
\end{figure}

In Fig.~\ref{Fig:cotro} we plot the expectation value of the
Polyakov loop in the pseudocritical range as a function of
temperature, for several values of the strength of the coupling in
the vector channel, at $\mu_5 = 300$ MeV. At $G_V=0$, this value
of $\mu_5$ is slightly above $\mu_5^c$. In the figure, solid line
corresponds to $G_V=0$; dashed line corresponds to $G_V = 0.25 G$;
dot-dashed line corresponds to $G_V = 0.5 G$; dotted line
corresponds to $G_V = 0.75 G$; finally, short dashed line
corresponds to $G_V = G$. As expected, the role of the vector
interaction is to smooth the phase transition in comparison with
the latter at $G_V = 0$. Indeed, the first order phase transition
at $\mu_5 = 300$ MeV at $G_V = 0$ is turned by the vector
interaction into a crossover at the same value of $\mu_5$.

\begin{figure}[t!]
\begin{center}
\includegraphics[width=8.5cm]{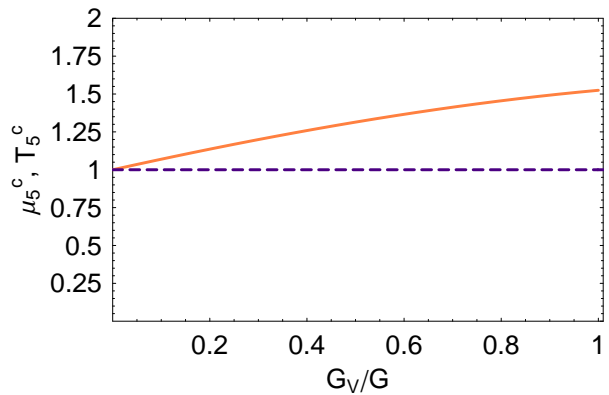}
\caption{({\em Color online}). Critical endpoint coordinates as a
function of the ratio $G_V/G$. Indigo dashed line corresponds to
the critical temperature; orange solid line corresponds to the
critical chiral chemical potential. Axial chemical potential and
temperature are measured in units of the endpoint coordinates at
$m_0 = m_{0,\text{Phys}}$.} \label{Fig:tatoeba}
\end{center}
\end{figure}

In Fig.~\ref{Fig:tatoeba} we plot the critical endpoint
coordinates, $\mu_5^c$ and $T^c$, as a function of the ratio
$G_V/G$. Blue dashed line corresponds to the critical temperature;
red solid line corresponds to the critical chiral chemical
potential. We find that the temperature of the critical endpoint
is poorly affected by the value of $G_V$ in the range that we have
examined; on the other hand, the critical value of the chiral
chemical potential is quite sensitive to the vector coupling.

It is useful to comment about the similarities between the phase structure at finite $\mu_5$,
which turns out from our computation, and the one well established (within effective models)
at finite baryon chemical potential, $\mu$. 
In this article, we have considered only one type of condensate, namely the chiral condensate,
which characterizes the symmetry breaking pattern at finite $T$ and $\mu_5$ in the fermion sector
of our model. Besides a smaller curvature of the critical line in the case of the chiral 
chemical potential, compared with that at finite $\mu$, we do not find qualitative
difference in the phase structure. Also, the effect of the vector interaction,
as well as of a finite current quark mass, is very similar, qualitatively, 
in the cases of finite $\mu$ and finite $\mu_5$. 
Some difference between the two phase diagrams might arise from different types of condensates.
For example, an important point that we have not considered in our article, for the sake of simplicity,
is the introduction of a diquark condensate, which might appear at finite mu5.
For the case of QCD at finite $\mu$, such a condensate is expected to be developed
at very large $\mu$~\cite{Alford:1997zt,Rapp:1997zu,Alford:1998mk}.
However, this possibility deserves further study and needs to be checked numerically by
dynamical computation of the condensates. We leave this point to a future project.

\section{Topological susceptibility}
In this Section we compute the topological susceptibility at
finite temperature, in presence of a background chiral density.
Topological susceptibility in QCD is defined as the correlator of
the topological charge at zero momentum; it can be computed from
the QCD partition function via the relation
\begin{equation}
\chi =
\left.\frac{\partial^2\Omega}{\partial\theta^2}\right|_{\theta=0}~,
\label{eq:poiu}
\end{equation}
where the $\theta$-angle is introduced by adding the following
term to the lagrangian density:
\begin{equation}
\theta\frac{g^2}{64\pi^2}\varepsilon^{\mu\nu\rho\sigma}F_{\mu\nu}^a
F_{\rho\sigma}^a~. \label{eq:ano}
\end{equation}
It is well known that one can get rid of the term in
Eq.~\eqref{eq:ano} in the QCD action by means of a chiral
rotation. However, after the chiral transformation, the $\theta$
dependence of the QCD partition function appears explicitely 
in the quark part of the QCD action. A model study having
some overlap with our study can be found
in~\cite{Sakai:2011gs}, where the full topological susceptibility
in presence of a background of topological charge is computed.
In the case of~\cite{Sakai:2011gs}, the topological charge
is introduced by adding the $\theta-$term to the lagrangian density,
which in turns acts as a source for the topological charge. 
The relevance of a finite $\theta$ in~\cite{Sakai:2011gs} lies on the
possibility that for temperatures higher than $\Lambda_{QCD}$, sphaleron transitions
induce a spacetime dependent $\theta$-angle~\cite{McLerran:1990de}. 

A low energy relation connects the vacuum chiral condensate to the
topological susceptibilty; in the theory of $N_f$ light flavors
the relation reads~\cite{Veneziano:1979ec,Leutwyler:1992yt}
\begin{equation}
\chi = |\langle\bar q q\rangle|
\left(\sum_f\frac{1}{m_f}\right)^{-1} +
O\left(\frac{m_f}{\Lambda_{QCD}}\right)~, \label{eq:Vene}
\end{equation}
where $\langle\bar q q\rangle$ is the common value of the quark
condensates for the light quarks. Equation~\eqref{eq:Vene} shows
that the topological susceptibility is proportional to the product
of the light flavor masses; thus, in the equation it is manifest
the fact that in a theory in which at least one massless flavor
exists, one has $\chi=0$. This was proved also by Ward identities
in~\cite{Crewther:1977ce}.

As stated before, a chiral rotation transmits the
$\theta$-dependence of the QCD action to the quark sector. After
the rotation, the quarks acquire a complex mass term. Besides, the
condensates that are not invariant under the axial rotation are
mixed among themselves. This scenario can be implemented within
the effective models to compute $\theta-$dependent quantities, see
for example~\cite{Fujihara:2005wk}. Beside this, the $\theta$-dependent action can
be used to compute $\chi$ by virtue of Eq.~\eqref{eq:poiu}. A
detailed discussion about the implementation of the
$\theta-$dependent lagrangian within the chiral model with the
Polyakov loop can be found in~\cite{Sakai:2011gs,Kouno:2011vu}. In
this article it is enough to mention that, in order to introduce
the $\theta$ angle in the model and compute the topological
susceptibility, it is enough to change $M=m_0 - 2\sigma G$ into the
grand potential with
\begin{equation}
\sqrt{\left[m_0\cos\left(\frac{\theta}{2}\right) -2\sigma G\right]^2
+ m_0^2\sin^2\left(\frac{\theta}{2}\right)}~. \label{eq:change}
\end{equation}

It is worth to mention here that generally speaking, at finite
$\theta$ other condensates might develop (and indeed they do
develop, see~\cite{Kouno:2011vu,Sakai:2011gs}). The presence of
these condensates makes the substitution~\eqref{eq:change} not
sufficient for the complete treatment of the $\theta$ angle within
the chiral models. However, such new condensates vanish in the
theory at $\theta=0$, to which we are interested; neglecting them
from the beginning does not change the derivative of $\Omega$ at
$\theta=0$. Therefore, the replacement in Eq.~\eqref{eq:change} is
sufficient for the purpose of computing the topological
susceptibility within this model.

\begin{figure}[t!]
\begin{center}
\includegraphics[width=8.5cm]{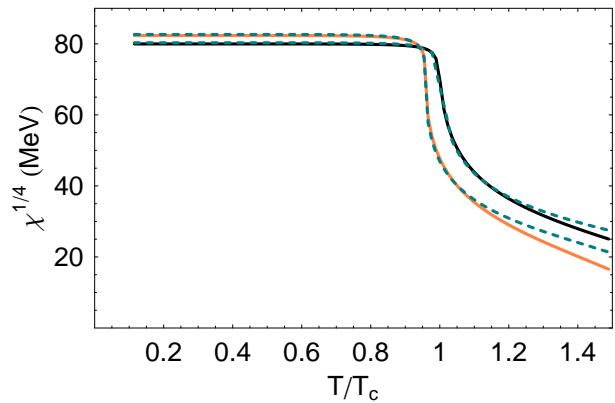}
\caption{Topological susceptibility as a function of temperature,
for several values of $\mu_5$. Black data correspond to $\mu_5 =
0$; orange data correspond to $\mu_5 = 300$ MeV. The green dotted
lines correspond to the right hand side of Eq.~\eqref{eq:Vene}.}
\label{Fig:chi}
\end{center}
\end{figure}

In Fig.~\ref{Fig:chi} we plot the topological susceptibility as a
function of temperature, for several values of $\mu_5$. At zero
temperature and chiral chemical potential we find
\begin{equation}
\chi = (79.97~\text{MeV})^4~,   \label{eq:R}
\end{equation}
which is in agreement with the large $N_c$ prediction in
Eq.~\eqref{eq:Vene}, which gives $\chi = (80.3~\text{MeV})^4$ with
two light flavors at $T=\mu_5 = 0$.

The relation~\eqref{eq:Vene} is actually satisfied within the
chiral model also at $T=0$ and $\mu_5\neq 0$. Indeed, a
straightforward computation shows that
\begin{equation}
\chi = -\frac{m_0}{2}\frac{N_c N_f}{2\pi^2}\sigma  {\cal I}~,
\label{eq:15}
\end{equation}
where
\begin{equation}
{\cal I} = G\int_0^\Lambda
\left[\frac{p^2 dp}{\sqrt{(p+\mu_5)^2+(m_0-2\sigma G)}} + \mu_5
\rightarrow -\mu_5\right]~;
\end{equation}
moreover, from the conditions $\partial\Omega/\partial\sigma=0$
and $\sigma = 2 \langle\bar uu\rangle$ we have
\begin{equation}
\langle \bar uu \rangle = - \frac{N_c N_f}{2\pi^2}\sigma  {\cal I} +
O\left(\frac{m_0}{m_q}\right)~. \label{eq:17}
\end{equation}
Here $m_q$ corresponds to the constituent quark mass. A comparison
between Eqs.~\eqref{eq:15} and~\eqref{eq:17} leads to
\begin{equation}
\chi = -\frac{m_0}{2}\langle \bar uu \rangle +
O\left(\frac{m_0}{m_q}\right)~,
\end{equation}
which is in agreement with Eq.~\eqref{eq:Vene} when the latter is
computed for two degenerate flavors. On the other hand, at finite
temperature we measure some discrepancy between the theoretical
prediction given by Eq.~\eqref{eq:Vene}, corresponding to the
green dotted line in Fig.~\ref{Fig:chi}, and the numerical results
obtained within the PNJL model (solid lines in the same figure).

\section{Conclusions}
In this article, we have reported our new results about the structure 
of hot quark matter in a background of a chiral density, the latter induced
by a finite axial chemical potential $\mu_5$. 
Firstly, we have analyzed analytically the effect of the axial chemical potential
on the chiral condensate, and on the critical temperature for restoration
of chiral symmetry. We restricted ourselves to the case of the NJL model,
and to the chiral limit. However, even within these simplifications, we are
able to understand the shape of the critical line at finite $\mu_5$.

Secondly, we have discussed the phase diagram
in the canonical ensemble formulation, in which $\mu_5$ is replaced by the
chiral density, $n_5$. This might be interesting because $n_5$ in QCD is connected
locally to the topological charge density, because of the integrated Ward identity.
As a consequence, it might be of interest to predict the numerical value
of the topological charge density along the critical line, as well as at the
critical endpoint. 

Thirdly, we have computed the effect of the vector interaction, as well as of the finite current quark mass, 
on the location of the critical endpoint. Such aspects should be taken
into account if a comparison with the Lattice data is desirable, and extend
the study started in~\cite{Ruggieri:2011xc}.  

As a final investigation, we have computed the full topological susceptibility
(i.e., which takes into account both the pure gauge and the dynamical fermion contribution)
at zero as well as at finite $\mu_5$, as a function of temperature. 
We find that the Di Vecchia-Leutwyler-Smilga-Veneziano (DLSV)
relation~\cite{Veneziano:1979ec,Leutwyler:1992yt} is satisfied at finite $\mu_5$
in the confinement phase of the model. At large temperature, above the critical temperature,
we measure a deviation from the DLSV relation, both at zero and at finite $\mu_5$;
this can be understood within the model, since terms 
of the order of $m_0/m_q$ which are negligible in the confinement phase, 
become important in the quark-gluon-plasma phase.

It is interesting to ask wether the work presented here can be improved. 
In our opinion, several directions are possible for future research.
As a first step, it would be important to investigate from a theoretical point
of view, how to couple correctly $\mu_5$ to a conserved quantity in QCD. Along this line, 
the work in Ref.~\cite{Rubakov:2010qi} seems quite enlightening. 
Secondly, it is of interest to include the possibility of other condensates,
among them a diquark condensate, in analogy to the situation of QCD at very large density.
We plan to report on these, as well as related, topics in the next future.

\acknowledgements We acknowledge stimulating discussions with
P. de Forcrand, M. D'Elia, G. Endrodi, M. Frasca, K. Fukushima,  
F. Negro,  O. Philipsen,  about the topics 
discussed in this article. Moreover, we thank T. Z. Nakano, A. Ohnishi,
Y. Sakai, T. Sasaki and H. Ueda for their careful reading of the manuscript
and their useful comments. M. R. also acknowledges the 
Japanese Society for the Promotion of Science for support during his staying
at Kyoto University, where par of this work was completed.

\end{document}